\documentclass{raa}           
\usepackage{graphicx,times}
\usepackage{natbib}
\usepackage{amssymb,amsmath}
\usepackage[T1]{fontenc}
\usepackage{ae,aecompl}
\usepackage{multirow}   
\usepackage{gensymb}    
\bibpunct{(}{)}{;}{a}{}{,}

\newcommand{\HI}{\hbox{\rmfamily H\,{\textsc i}}}
\newcommand{\HIsub}{\hbox{{\scriptsize H}\,{\tiny I}}} 

\newcommand{\Msun}{\hbox{$\rm M_\odot$}}

\usepackage[a4paper=true,pagebackref=true]{hyperref}
\hypersetup{colorlinks = true, linkcolor = green, anchorcolor = red, citecolor = blue, filecolor = red, pagecolor = red, urlcolor = red}

\begin{document}
\title{The conformity of HI galaxies in ALFALFA-SDSS sample}
 \volnopage{ {\bf 20XX} Vol.\ {\bf X} No. {\bf XX}, 000--000}
   \setcounter{page}{1}

\author{Lincheng Li\inst{1,2,3}, 
   Bo Qin\inst{1}, 
   Jie Wang\inst{1,2},
   Jing Wang\inst{4},
   Yougang Wang\inst{1}
   }
\institute{National Astronomical Observatories, Chinese Academy of Sciences, Beijing 100012, China; {\it lilincheng@bao.ac.cn}\\
\and
School of Astronomy and Space Science, University of Chinese Academy of Sciences, Beijing 100049, China\\
\and
International Centre for Radio Astronomy Research (ICRAR), The University of Western Australia, 35 Stirling Hwy, Crawley, WA, 6009, Australia\\
\and 
Department of Astronomy, Peking University, Beijing 100871, China\\ 
\vs \no
   {\small Received 20XX Month Day; accepted 20XX Month Day}
}

\abstract{The conformity effect, indicating the evolution of a galaxy is related to its surrounding neighbour galaxies as far as a few Mpc, is an interesting phenomenon in the modeling of galaxy and evolution. Here we study the conformity effect of HI galaxies in a matched galaxy sample between SDSS DR7 and ALFALFA surveys. By checking the probability difference for the detected $\HI$ galaxies as a function of distance around a normal or an $\HI$ galaxy, we find that this effect is significant out to 5 Mpc. It also shows a dependence on the stellar mass of galaxies, with the strength the strongest in the stellar mass range of $10^{10}$<$M_*$/$\Msun$<$10^{10.5}$. 
However, when the sample is confined to central galaxies in groups with virial radii smaller than 1 Mpc, the 1-halo conformity signal is still evident, while the 2-halo conformity signal is reduced to a very weak amplitude. Our results confirm the previous study in the optical band that the 2-halo term is possibly caused by the bias effect in the sample selection. Furthermore, we confirm the existence of the 1-halo conformity discovered in the optical and radio band in previous lectures. Our results provide another viewpoint of the conformity effect and hope to shed the light on the co-evolution of the galaxies and their neighbours in the current galaxy formation models. 
\keywords{method: data analysis, radio lines: galaxies, galaxies: evolution
}
}

\authorrunning{L. Li et al.} 
\titlerunning{The conformity in ALFALFA-SDSS sample}  
\maketitle

%
\section{Introduction} 
\label{sect:intro} 
In recent years, an intriguing clue regarding the properties of galaxies is the 'galactic conformity', a controversial phenomenon that the properties of galaxies are correlated with their more massive central galaxies with star-forming centers having preferentially more star-forming satellites. This intra-halo phenomenon was originally discovered by \citet{Weinmann06}, and was confirmed by several follow-up studies (\citealt{Kauffmann10}; \citealt{Knobel15}; \citealt{Phillips14}).
Remarkably, \citet{Kauffmann13} and \citet{Kauffmann15} have shown that the conformity effect on specific star formation rate persists out to several Mpc, many times larger than the virial radii of the dark matter halos concerned.
Generally, the intra-halo and inter-halo conformity effects have been dubbed '1-halo conformity' and '2-halo conformity' respectively. 

While the 1-halo conformity could presumably be explained by group-scale physical processes \citep{Hearin15}, the 2-halo conformity is however challenging the traditional halo occupation model of galaxies, in which the statistical properties of galaxies, such as the luminosity function, stellar mass function and spatial clustering of galaxies, can be solely determined by the halo mass (see \citealt{Cooray02} for a review). 
Hence, many recent efforts have been devoted to developing models of galactic conformity (e.g.\citealt{Lu15}; \citealt{Hearin15}; \citealt{Hearin16}; \citealt{Henriques16}; \citealt{Pahwa17};). In these models, some exotic effects such as the preheating of the gas by energy input from AGNs and the correlation of halo assembly history with the environment or 'halo assembly bias', might explain the observed data.  

One important issue for these models is that on which scale and to what extend do the 2-halo conformity exists.  \citet{Calderon18} investigated the conformity signal in Sloan Digital Sky Survey (SDSS) DR7 on the colour, specific star formation rate and morphology of galaxies. Their results with marked correlation methods reveal a small, yet highly significant signal for all three properties in low mass groups and scales of 0.8$\sim$4$h^{-1}$ Mpc. With the stellar mass and local density, \citet{Sin19} developed a conceptual framework and methodology to explore hidden variables which control the quenching of galaxies by investigating the residual, $\Delta$, between the observed quenched fraction of galaxies and the predicted value. After applying the analysis to a local galaxies sample, they found $\Delta$ is correlated out to 3 Mpc, suggesting that halo-related properties need to be considered for galaxy quenching. But the significance of their result is unclear and relies on the accuracy of the group catalogue they used.

However, recent studies showed that the interpretation of the conformity signal strongly depends on the methodology and the data used in the analysis. \citet{Sin17} pointed out that the strong large-scale conformity signal in \citet{Kauffmann13} was a result of the combination of three bias factors in their analyse: the unequal weighting towards the over-dense regions, the misclassification of centrals and the use of median to describe the bimodal distribution of sSFR. The conformity signal reduced significantly after they made modifications to these issues.
Similarly, \citet{Tinker18} and \citet{Sun18} conclude that the significant 2-halo conformity detected in \citet{Kauffmann13} originates from some satellite galaxies misclassified as centrals and no new physical processes are needed to reproduce the observed results of conformity in simulations.
\citet{Zu18} also showed that their fiducial halo mass quenching model has been able to successfully explain the overall environmental dependence and the conformity of galaxy colours in Sloan Digital Sky Survey, without any galaxy assembly bias.

These results in the optical band make it murky whether the 2-halo conformity indeed exists. 
A parallel investigation of 2-halo conformity in radio band is promising to provide a double-check and more insights into the physical origins of it.  
As the pioneering work in the $\HI$ band, \citet{Enci15} utilise the data cubes of Bluedisk project and found that that the companions around $\HI$-rich galaxies tend to $\HI$-rich as well.
With the same data cubes, \citet{Jing15} also shows that galaxies whose $\HI$ mass function is high relative to the standard scaling relations have an excess $\HI$ mass in the surrounding environment as well, suggesting a real conformity effect.
These works have demonstrated the feasibility of investigating galactic conformity through the radio window. 
However, limited by the data set, they did not check the conformity for HI-poor galaxy population and on scales larger than the virial radius of a normal halo yet. A larger data set with more galaxy sample and a continuous sky coverage is needed to figure out how the strength of galactic conformity changes with stellar mass and the physical scale.

In this paper, we combine SDSS DR7 with $\alpha.$70 (70 percent data of ALFALFA survey and the group catalog of \citet{Yang07} to perform a statistical analysis on the scale and stellar mass dependence of the 2-halo conformity. Meanwhile, if the conformity signal can be detected with the HI data, we aim to check whether the conformity signal can be explained by the bias effects found in the optical band. 

The paper is organised as follows: in Section 2 we introduce the data sets used in the analyse. The results are presented in Section 3. We compare our results with previous studies in Section 4 and summarise in Section 5.


\section{Data} 
\subsection{The Galaxy Samples}
\label{sec:opt sample} 
The $\HI$ data that we use are drawn from the Arecibo Legacy Fast ALFA (ALFALFA) survey \citep{ALFA70}, $\alpha.70$.
   The ALFALFA survey is a blind extra-galactic $\HI$ survey exploiting the Arecibo telescope to conduct a census of the local $\HI$ universe over $\sim$ 7000 deg$^2$ of the  high Galactic latitude sky visible from Arecibo out to $z\sim~0.06$. The minimum $\HI$ mass it detected reaches $4.4\times10^6 \Msun$ at 10 Mpc for 5$\sigma$ detection with a velocity resolution of 30 km/s.
   The data release, $\alpha.70$, covering its whole redshift range, contains 70$\%$ of its data and includes 25535 detections. 
   
\begin{figure} 
\includegraphics[width=0.7\columnwidth,angle=270]{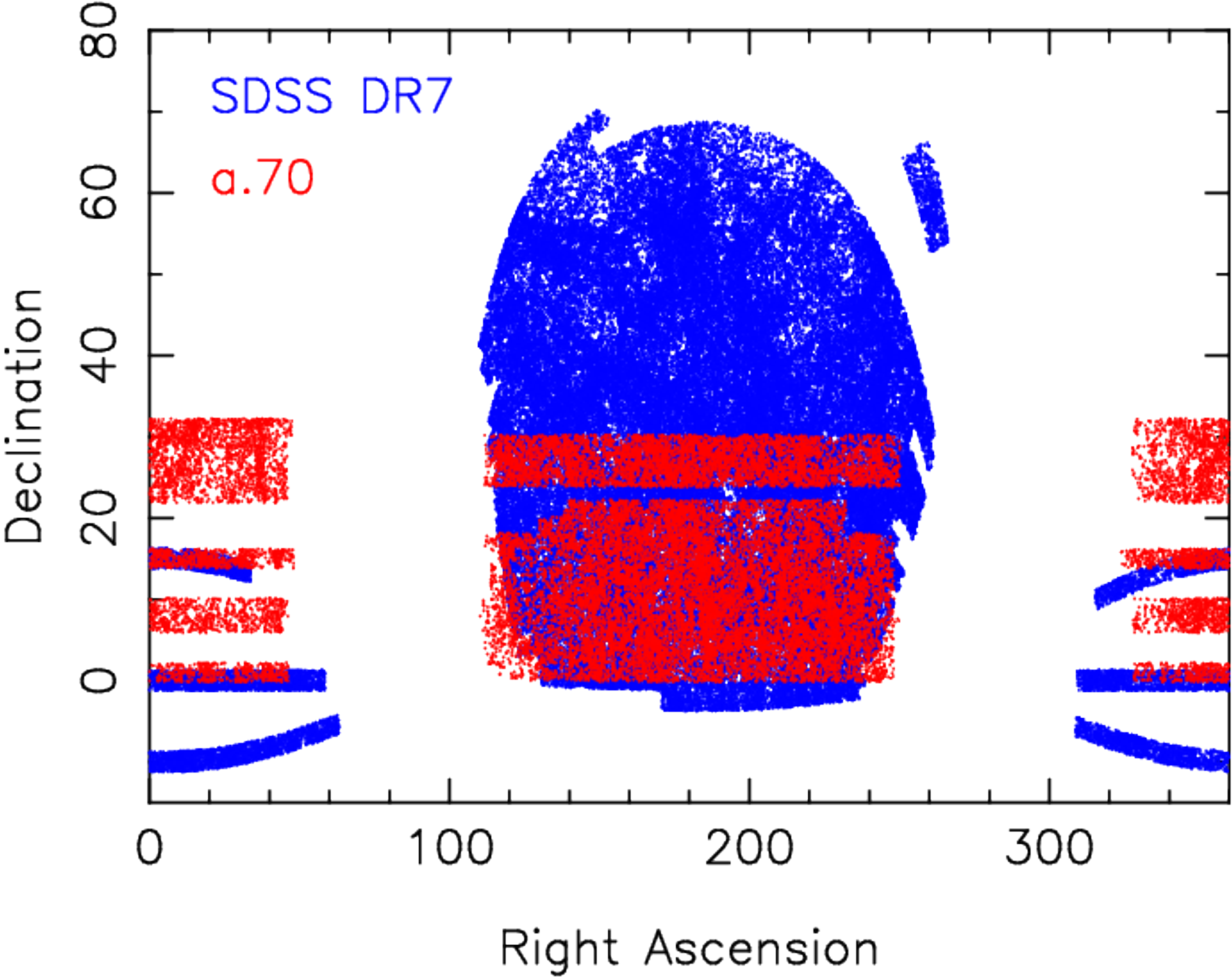}
\caption{The footprints of $\alpha.70$ and SDSS DR7. The blue and red regions are for $\alpha.70$ and SDSS DR7, respectively. }
\label{fig_footprint}
\end{figure}

   In Fig.\ref{fig_footprint} we show the footprints of SDSS DR7 (blue dots) and $\alpha.70$ (red dots). As can be seen, a large fraction of the footprint of $\alpha.70$ overlaps with that of the SDSS DR7. The majority ($77\%$) of the detections in $\alpha.70$ are  highly reliable with an $S/N>6.5$ and the left are relatively lower signal-to-noise detections but they are also likely to be real detections. As the sample is analyzed in a statistical way, we include all these $\HI$ detections with optical counterparts in our study.
 
\subsection{Cross-matching sample}  
$\alpha.70$ provides a good $\HI$ data-set to study the atomic gas content in galaxies. However, a cross-matching between the $\alpha.70$ and optical catalogs is needed before any further analysis.   
The ALFALFA team has made a cross-match between the $\alpha.70$ detections and the SDSS, and provided the coordinates of the optical counterparts (OC). In total, 23881($94\%$) $\HI$ detections have assigned OCs. Most of the remaining 1654 detections have negative redshifts and are classified as High Velocity Clouds (HVC).  

In order to cross-match between the group catalog of \citet{Yang07} and the catalog of stellar mass, we obtained the IDs in the SDSS data-set by searching the nearest primary objects as their OC in SDSS DR7 within a radius of 1 arcmin on the SDSS DR7 SkyServer website (http://cas.sdss.org/dr7/en/tools/crossid/crossid.asp). In total, 20438 objects are matched. The other $\HI$ detections lie outside the footprint of SDSS DR7, so no cross-matching can be performed. 
   
To test the reliability of our matched result, we compare our result with $\alpha.40$, the 40$\%$ of ALFALFA data, among which the $\HI$ detections have been matched with the SDSS DR7 data. The match of $\alpha.40$ was processed by the ALFALFA team in several years with his/her scientific judgment based on objects' optical information so we treat it as a reliable standard. 
A comparison between our result and $\alpha.40$ shows that for the galaxies included in both $\alpha.70$ and $\alpha.40$, more than 95$\%$ objects in our matched result are consistent with the matched result of $\alpha.40$ 
-- suggesting that our matched result is reliable for further statistical analysis.
   
The ALFALFA $\HI$ centroid position accuracy was to about 18 arcsec. 
The coordinates and redshifts of the $\HI$ detections in $\alpha.70$ have slight offsets from their OCs. We use the coordinates and redshifts of OCs when computing the distance in the following sections.  

\subsection{Definition of HI-rich, HI-poor and normal galaxies}  
\label{sec:define_HI} 
We now define $\HI$-rich/$\HI$-poor galaxies which are indeed redshift related because ALFALFA was flux-limited as a blind survey. We classify the galaxies in our sample into three categories: $\HI$-poor, normal and $\HI$-rich in two steps -- first we divide the redshift coverage into different bins of 500 km/s each. Second, the stellar mass range is divided into different Log$M_*/\Msun$ bins of 0.25 each. 
In each redshift and stellar mass bin, galaxies with the top (bottom) 30 percent $M_{\HIsub}$ are classified as $\HI$-rich ($\HI$-poor) galaxies, and the rest are classified as normal galaxies.
   
\begin{figure} 
\centering
\includegraphics[width=0.7\columnwidth,angle=270]{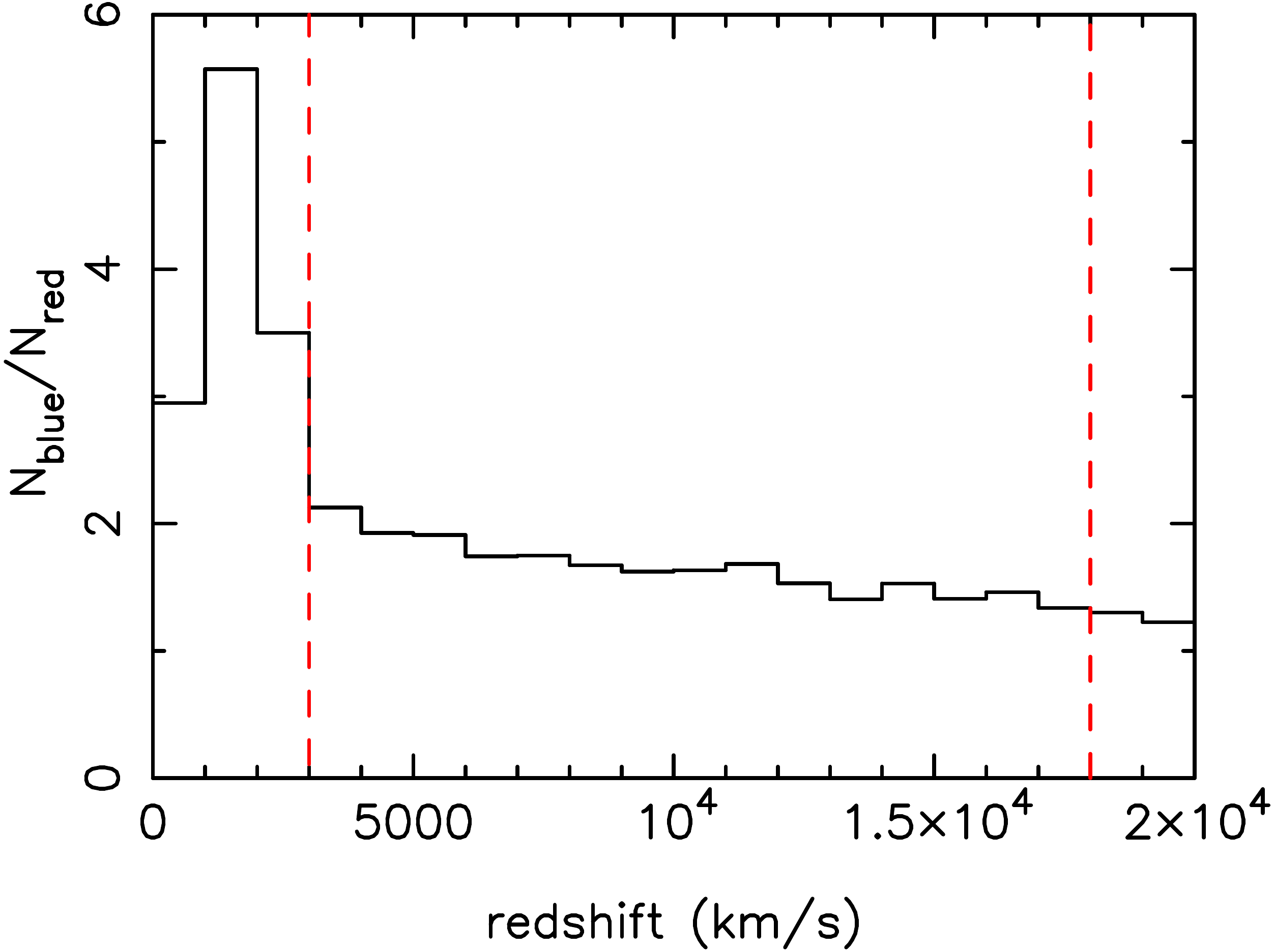}
\caption{The ratio of $N_\text{blue}/N_\text{red}$ in SDSS DR7 as a function of redshift, where $N_\text{blue}$ and $N_\text{red}$ are the numbers of red galaxies and blue galaxies in each redshift bin. Two red dashed lines indicate the redshift range of our $\HI$-detected sample: 3000 km/s $\sim$ 18000 km/s.}
\label{fig_brratio}
\end{figure}

 For a flux limited sample, like SDSS, due to the limiting magnitude of the sample and the smaller mass-to-light ratio of blue galaxies (thus brighter at fixed mass), it is easier for a blue galaxy at low redshift to be observed than a red galaxy of similar mass. Because of this selection effect, the optical galaxy sample is biased towards blue galaxies in SDSS, especially at low redshifts. In Fig.\ref{fig_brratio} we show the result of $N_\text{blue}/N_\text{red}$, defined as the number of blue galaxies over the number of red galaxies (the definition of blue and red galaxies in \citealt{Yang07} is used), as a function of redshift. As can be seen, this bias on color turns significant when $cz$ is less than 3000 km/s.
Therefore, we excluded the galaxies with $cz<3000$ km/s in our study. 
In total, our sample includes 131076 optical galaxies among which 14173 have been detected by ALFALFA in the redshift range we analyze (3000$\sim$18000 km/s).

\section{Results}  
\subsection{Galactic conformity} 
\begin{figure} 
\includegraphics[width=0.7\columnwidth,angle=270]{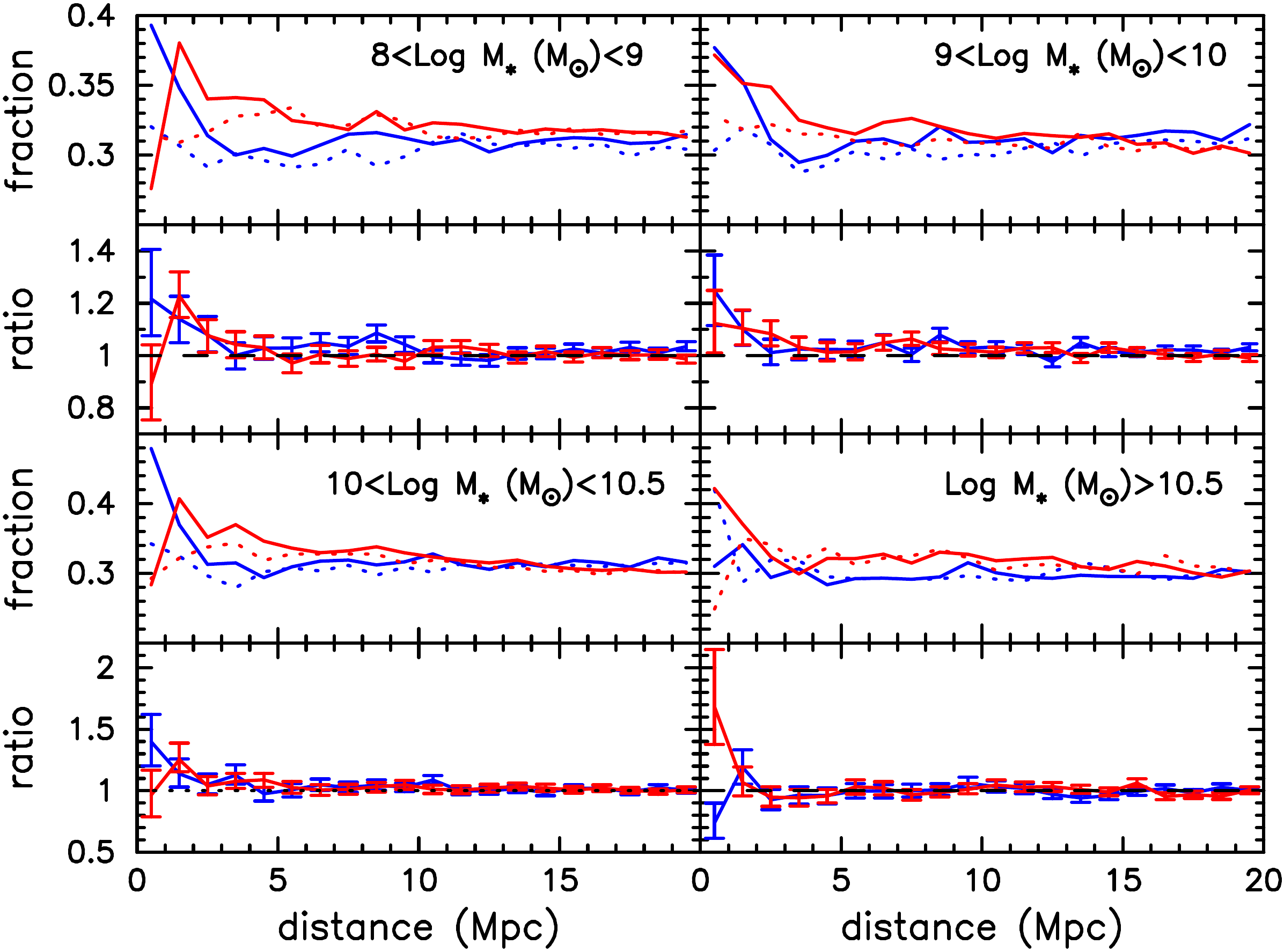}
\caption{The fractions of $\HI$-rich and $\HI$-poor neighbors as a function of distance for $\HI$-rich, $\HI$-poor and normal galaxies. 
Blue (red) solid lines indicate the fraction of $\HI$-rich ($\HI$-poor) neighbors of $\HI$-rich ($\HI$-poor) galaxies, $f_\text{rr}$ ($f_\text{pp}$).
Blue (red) dotted lines represent the fraction of $\HI$-rich ($\HI$-poor) neighbors for normal galaxies, $f_\text{rn}$ ($f_\text{pn}$).
The ratios of the values between solid and dotted lines in each color are plotted at the bottom of each panel with the same color. 
Error bars show the 80$\%$ confidence intervals from bootstrap resampling.}
\label{fig_rprp}
\end{figure}

By the definitions of $\HI$-poor/$\HI$-rich and normal galaxies in \ref{sec:define_HI}, we compute\\
(1) the fraction of $\HI$-rich galaxies around $\HI$-rich galaxies, $f_\text{rr}=N_{\text{rich}}/N_{\text{tot}}$,\\
(2) the fraction of $\HI$-poor galaxies around $\HI$-poor galaxies, $f_\text{pp} = N_{\text{poor}}/N_{\text{tot}}$,\\
(3) the fraction of $\HI$-rich galaxies around normal galaxies, $f_\text{rn}=N_{\text{rich}}/N_{\text{tot}}$,\\
(4) the fraction of $\HI$-poor galaxies around normal galaxies, $f_\text{pn}=N_{\text{poor}}/N_{\text{tot}}$\\ 
as functions of the distance to the $\HI$-poor/$\HI$-rich/normal galaxies, where $N_{\text{rich}}, N_{\text{poor}}$ and $N_{\text{tot}}$ are the number of $\HI$-rich neighbours, the number of $\HI$-poor neighbours and the total number of neighbours around primary galaxies, respectively.
Fig.~\ref{fig_rprp} shows the results in four stellar mass ranges of the $\HI$-poor/$\HI$-rich/normal galaxies. 
The blue solid, red solid, blue dotted and red dotted lines indicate the results of $f_\text{rr}$, $f_\text{pp}$, $f_\text{rn}$ and $f_\text{pn}$, respectively. 
Error bars are the 80 confidence level computed with bootstrap resampling. 
At the bottom of each panel, the ratios of $f_\text{rr}$-to-$f_\text{rn}$ and $f_\text{pp}$-to-$f_\text{pn}$ are plotted in blue and red solid lines. The horizontal black dashed lines denote the fiducial level of 1.

As shown in Fig.~\ref{fig_rprp}, for galaxies with $8<$Log$M_*/\Msun<10.5$, the ratios of $f_\text{rr}$-to-$f_\text{rn}$ and $f_\text{pp}$-to-$f_\text{pn}$ are remarkably higher than 1 within 2 Mpc, and decline gradually with increasing distance, which indicates that more $\HI$-rich ($\HI$-poor) galaxies are found near $\HI$-rich ($\HI$-poor) galaxies compared to normal galaxies. 
This tendency is almost independent of the stellar mass but becomes much weaker in the range Log$M_*/\Msun>10.5$, and it seems to hold out to a few Mpc with the error bars however comparable with the excess of the ratios.

\begin{figure} 
\includegraphics[width=0.7\columnwidth,angle=270]{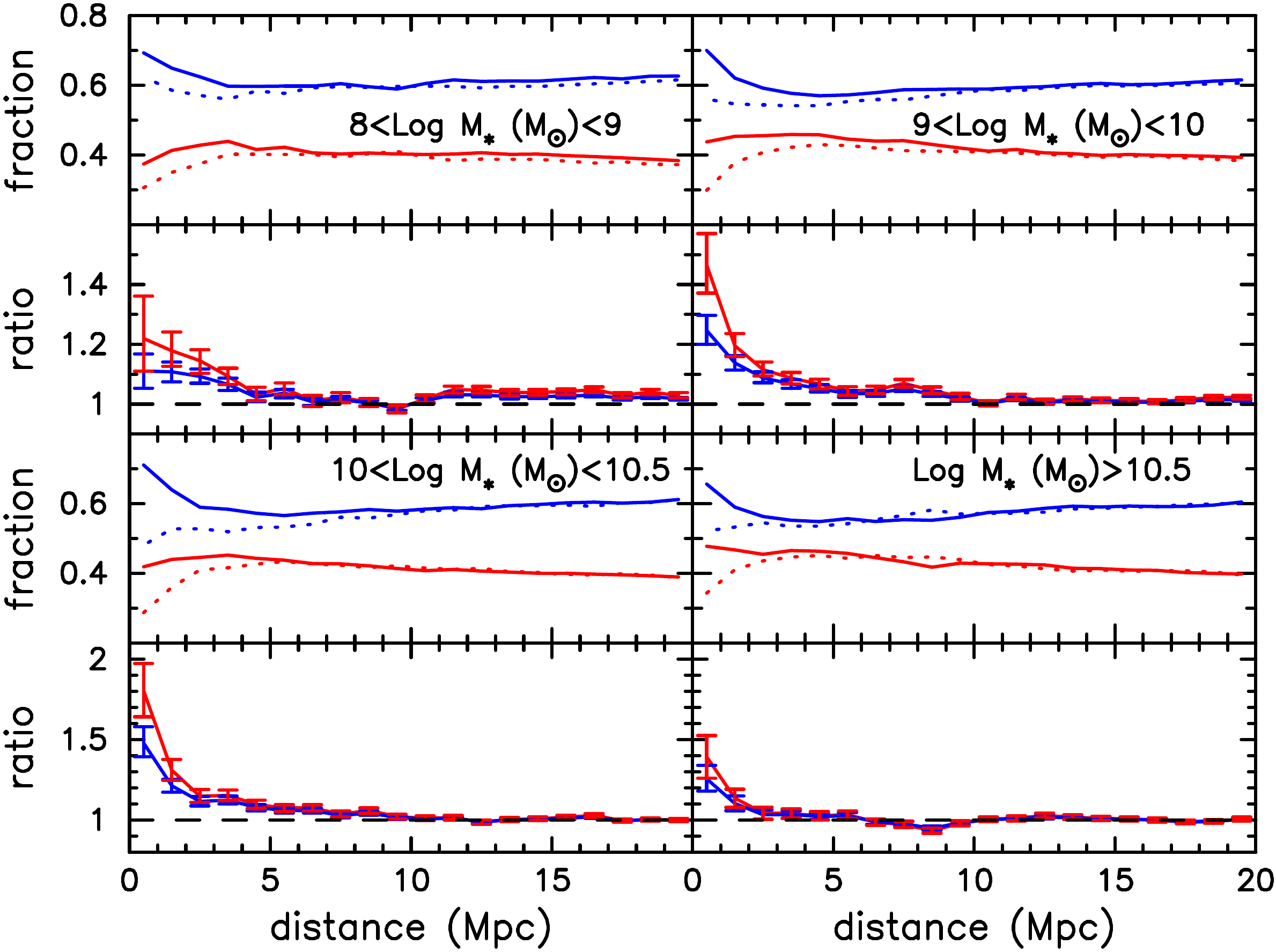}
\caption{The fractions of red and blue neighbors as a function of distance for $\HI$-poor, $\HI$-rich and normal galaxies.
Blue solid/dotted lines represent the fraction of blue neighbors for  $\HI$-rich/normal galaxies.
Red solid/dotted lines represent the fraction of red neighbors for $\HI$-poor/normal galaxies. 
The ratios of the values between solid and dotted lines in each color are plotted on the bottom of each panel with the same color. 
The black dashed lines indicate the fiducial value of 1. 
Error bars show the 80$\%$ confidence level from bootstrap resampling.}
\label{fig_brrp}
\end{figure}

In order to increase the signal-to-noise ratio, we still use the HI galaxies as the primary galaxy, but intend to find the neighbouring galaxies in larger optical sample -- the SDSS DR7 sample; or searching the neighbouring $\HI$ galaxies around the normal galaxies. 
Since the $\HI$ content of a galaxy is correlated with its color and star formation activities, we should also expect to find similar conformity signals in the excesses of the fractions of blue/red galaxies around $\HI$-rich/$\HI$-poor galaxies compared to normal galaxies. 
Therefore, we compute \\
(1) the fraction of red galaxies around $\HI$-poor galaxies, $f_\text{rp}=N_\text{red}/N_\text{tot}$; \\
(2) the fraction of blue galaxies around $\HI$-rich galaxies, $f_\text{br}=N_\text{blue}/N_\text{tot}$; \\
(3) the fraction of red galaxies around normal galaxies, $f_\text{ren}=N_\text{red}/N_\text{tot}$; \\
(4) the fraction of blue galaxies around normal galaxies, $f_\text{bn}=N_\text{blue}/N_\text{tot}$;\\
as functions of the distance to the $\HI$-poor/$\HI$-rich/normal galaxies,
where $N_{\text{red}}$ and $N_{\text{blue}}$ are the number of $\HI$-rich and $\HI$-poor neighbours around primary galaxies, respectively.
Fig.~\ref{fig_brrp} shows the results in four stellar mass ranges of the $\HI$-poor/$\HI$-rich/normal galaxies. The blue solid, red solid, blue dotted and red dotted lines indicate the results of $f_\text{br}$, $f_\text{rp}$, $f_\text{bn}$ and $f_\text{ren}$ respectively. 
At the bottom of each panel, the $f_\text{br}$-to-$f_\text{bn}$ ratios and the $f_\text{rp}$-to-$f_\text{ren}$ ratios are also plotted as blue and red solid lines.  
As can be seen, Fig.~\ref{fig_brrp} confirms the results of Fig.~\ref{fig_rprp} with smaller error bars and we can clearly see the conformity signal persists out to a few Mpc, much larger than the virial radius of a dark matter halo that a normal galaxy resides in.
In addition, this galactic conformity shows a clear dependence on the stellar mass of galaxiese, with the strength the strongest in the stellar mass range of $10^{10}$<$M_*$/$\Msun$<$10^{10.5}$.
The results found in Fig.~\ref{fig_rprp} and Fig.~\ref{fig_brrp} are qualitatively consistent with the conclusion of \citet{Kauffmann13}, in favor of the 2-halo conformity extending out to several Mpc. 

In our sample, even we excluded galaxies below $z=0.01$, the ratio of $N_{\rm blue}/N_{\rm red}$ still depends on the redshift weakly. But the signal of the conformity only extends to 5 Mpc, corresponding to a very shallow redshift region, and the signal is defined as the ratio between the fractions from the constrained sample and from the total sample, we do not think the above redshift dependence will affect our conclusions significantly. 

\begin{figure} 
\includegraphics[width=0.7\columnwidth,angle=270]{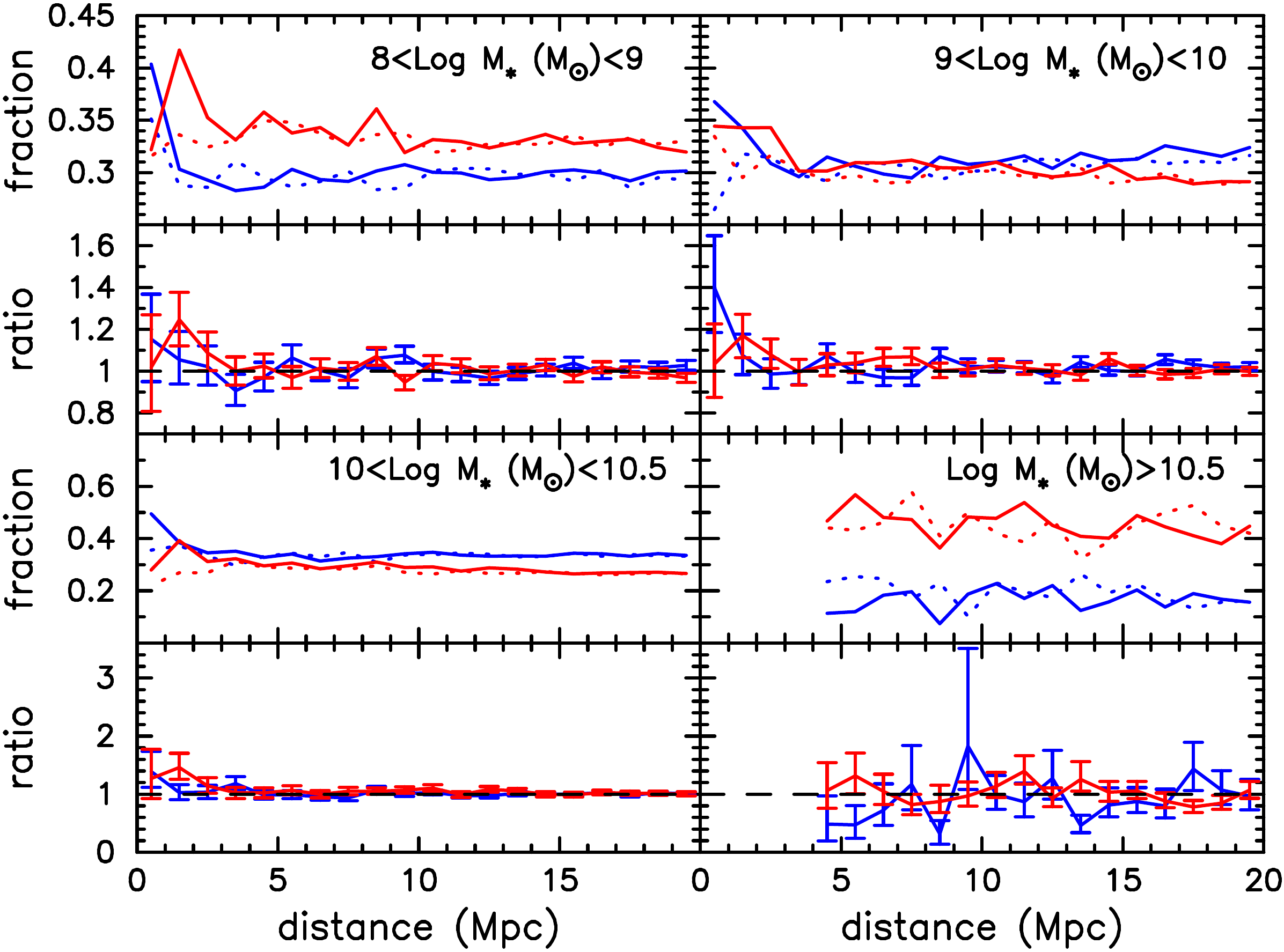}
\caption{Same as Fig.~\ref{fig_rprp} except that all satellites and centrals in halos with virial radii larger than 1 Mpc are excluded. The results within 4 Mpc in the most massive panel are vacant due to the lack of galaxy sample.}
\label{fig_rprpcr1}
\end{figure}

\begin{figure} 
\includegraphics[width=0.7\columnwidth,angle=270]{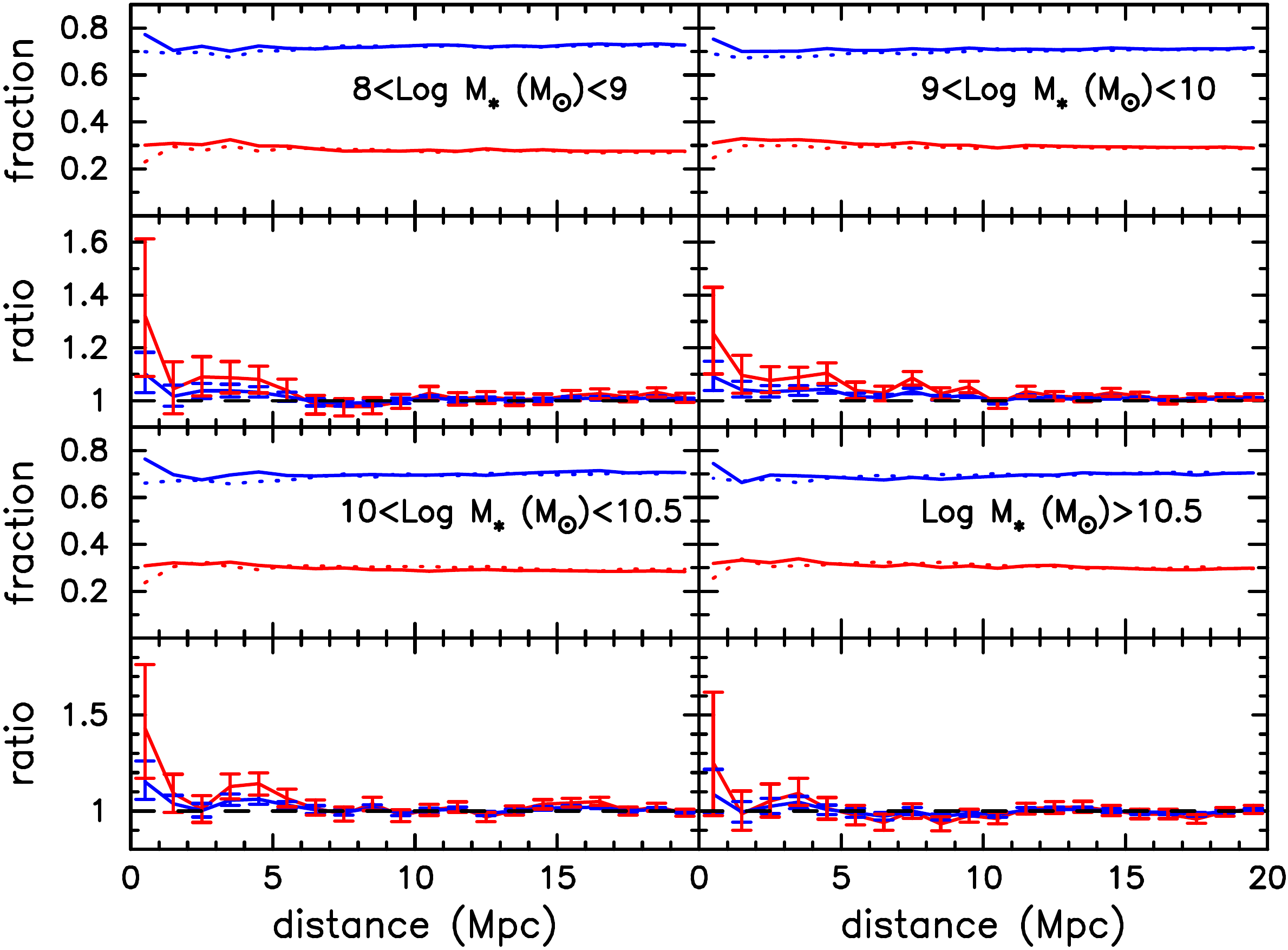}
\caption{Same as Fig.~\ref{fig_brrp} except that all satellites and centrals in halos with virial radii larger than 1 Mpc are excluded.}
\label{fig_brrpcr1}
\end{figure}

To find out whether the results above are also caused by the bias of the sample found by \citep{Sin17,Tinker18,Sun18,Zu18}, we followed \cite{Sin17} and excludeD all satellite galaxies as well as central galaxies in halos with virial radii larger than 1 Mpc (the virial radii are derived from the halo mass in Yang's catalogue) and compute the ratios in Fig.~\ref{fig_rprp} and Fig.~\ref{fig_brrp}.  
The results are showed in Fig.~\ref{fig_rprpcr1} and Fig.~\ref{fig_brrpcr1}, respectively.

With the removal of all satellites and centrals in halos with radii greater than 1 Mpc, the conformity signal in Fig.~\ref{fig_rprpcr1} still exists, but gets weaker and extends to a smaller scale.  This trend is more clear in Fig.~\ref{fig_brrpcr1} when the signal-to-noise ratio increased, especially for the bins of distance larger than 1 Mpc.
For the results with distance less than 1 Mpc, the ratios decrease compared to Fig.~\ref{fig_brrp} but remain at a remarkable level.
In addition, the conformity shows a dependence on the stellar mass, with the amplitude highest in the stellar mass range of $10^{10}<M_*<10^{10.5}\Msun$. This can be interpreted as the evidence of the 1-halo conformity. 
While for the distance larger than 1 Mpc, the conformity gets very weak, although the signal is more than 1 sigma significant between 3 Mpc and 6 Mpc on the distance for all stellar mass bins, which probably indicates that some mechanisms beyond halo mass are driving the connections of galaxies evolution on these scales. As what \citet{Sin17} did, we also tried to exclude the satellites within 4 Mpc, 2 Mpc and 1 Mpc, and confirmed that the signal of conformity gets weaker when smaller cutting scales are used. So we only show the result for the option of 1 Mpc here. 

\section{Comparison With previous results and Discussions}
\label{sec:discussion} 

Our results of simple statistics of the whole $\HI$ galaxy sample confirmed the conformity signal at large scale found in the optical band.  \citet{Kauffmann13} used a volume-limited sample of galaxies drawn from the SDSS DR7 with $M_*>2\times10^{11}\Msun$ to investigate the scale of conformity effect and how it changes as a function of the mass of the central galaxy. They showed the galactic conformity on the specific star formation rates and the pseudo-HI gas mass fractions of satellites. For centrals with 11<Log$M_*/\Msun$<11.5 the conformity effect is confined to scales less than 1-2 Mpc while for centrals with 10<Log$M_*/\Msun$<10.5, galactic conformity persists to 4 Mpc (the largest scale they concerned).  We also found the conformity extends to as far as 4 Mpc for the stellar mass bin 8<Log$M_*/\Msun$<10.5 and just go to 1-2 Mpc for the massive bin Log$M_*/\Msun$> 10.5. While the strength also gets weaker in the most massive bin. 

Furthermore, we confirmed in the radio band that the 2-halo conformity becomes weaker and extends to smaller scale when the sample is confined to centrals in halos with virial radius smaller than 1 Mpc, even in our cross-correlated sample between the radio and optical bands. This result agrees with the arguments raised in the optical band in \citep{Sin17, Tinker18, Sun18, Zu18}. 

In the radio band, \citet{Enci15} analysed the $\HI$ data cubes of the Bluedisk project and found that companions around $\HI$-rich galaxies tend to be $\HI$-rich as well. With the same data cubes, \citet{Jing15} found that the galaxies having high $\HI$ mass function also have an excess $\HI$ mass in the surroundings within a distance of 500 kpc. Our results agree well with these work given that the conformity signal always exists in the first bin of distance in our results. Therefore, we suggest that the one-halo conformity signal is true in the radio band as well. 

In this study we have not compared our results with any theoretical models, and are not sure if the current physical model is enough to explain the weak signal at scales beyond 1 Mpc in our measurements. Interestingly, \citet{Rafie18} investigated the $\HI$ conformity signal in the MUFASA hydrodynamical simulation. Although too many low-mass galaxies are quenched in the simulation, their result is in accord with our results that the 2-halo conformity signal declines more quickly with distance in massive halos ($M_\text{halo}$>$10^{12}\Msun$). Nevertheless, massive halos show a stronger conformity than less massive halos within 1 Mpc in their results, which is not consistent with our results. This inconsistency is probably due to the immature gas model in their simulation given that the fraction of quenched low-mass galaxies in MUFASA is too high compared to observations. We need to check that in further study with more reliable galaxy formation simulations.

 Another caveat for the reader is that the results found in this paper are based on the flux-limited sample from ALFALFA. Therefore the results actually are based on a HI detected sub sample matched to the SDSS optical sample. However, we changed the definition of HI rich/poor sample with different top/bottom fraction of the whole sample, the results do not change too much. This indicates that our results not so sensitive to the flux limit in this sample. But more further careful checks in a complete sample is needed in the future given the current complete sample is small for this check.

\section{Summary}
  In this paper, we have combined the optical galaxy sample from SDSS DR7 and the $\HI$ galaxy sample from $\alpha.70$ to investigate the conformity effect in radio band and estimate the scale on which the environment starts to affect the gas content in galaxies.  
 
 If we only measure the conformity signal in the whole matched sample, in the stellar mass range we considered here, the $\HI$ galactic conformity is found to persist out to $\sim$ 5 Mpc. This signal shows a dependence on the stellar mass of galaxies, with the strength the strongest in the stellar mass range of $10^{10}$<$M_*$/$\Msun$<$10^{10.5}$.
 
 We follow the consideration of the sample selection bias effect in the optical band \citep{Sin17}, and check if the 2-halo term of conformity could be affected by this effect in the radio band as well.  When all satellites and centrals in halos with virial radii greater than 1 Mpc are excluded from the sample, the 2-halo conformity signal is significantly reduced. We confirmed the arguments in the radio band that the 2-halo conformity signal is possibly caused by the bias effects in the sample. However, as \citet{Crain17} pointed out, it is not easy to predict the $\HI$ content in each galaxy for the current galaxy formation model. There is still a long way to compare the model prediction with our results and tell if the current model is enough to explain the two halo term conformity in the radio band. 

 In addition, we confirmed the finding in the Bluedisk project by \citet{Enci15, Jing15} with the 1-halo conformity signal in the matched sample between $\alpha.70$ and SDSS DR7. Further understanding of the 1-halo conformity in the radio band maybe could provide some hints on the physical processes involved in the star formation activities in a single halo's potential.


\normalem
\begin{acknowledgements}
We thank Li Shao and Ming Zhu for helpful discussions, and the referee's very useful comments and suggestions on improving the paper. This work was funded by the National Key R{\&}D Program of China under grant number 2018YFA0404603. The project was also supported by CAS Interdisciplinary Innovation Team (JCTD-2019-05).
\end{acknowledgements}
  
\bibliographystyle{raa}
\bibliography{references}

\end{document}